\documentclass[prl,twocolumn,showpacs]{revtex4}
\usepackage{graphicx}

\begin{document}

\title{Is Arnold diffusion relevant to global diffusion?}
\author{Seiichiro Honjo}
\email{honjo@complex.c.u-tokyo.ac.jp}
\author{Kunihiko Kaneko}
\affiliation{
Department of Basic Science, Graduate School of Arts and Sciences,
The University of Tokyo, 3-8-1 Komaba, Meguro, Tokyo 153-8902, Japan
}
\date{\today}

\begin{abstract}
Global diffusion of Hamiltonian dynamical systems is investigated by using a
coupled standard maps.  Arnold web is
visualized in the frequency space, using local
rotation numbers, while Arnold diffusion and resonance overlaps are 
distinguished by the residence time distributions at
resonance layers. Global diffusion in the phase space is shown to be
accelerated by diffusion across overlapped resonances generated by the coupling term,
rather than Arnold diffusion along the lower-order resonances.
The former plays roles of
hubs for transport in the phase space, and accelerate the diffusion.
\end{abstract}

\pacs{05.45.-a, 05.20.-y, 05.60.Cd}
\maketitle

Understanding in global dynamic behavior in Hamiltonian systems is a fundamental issue
in nonlinear dynamics and statistical physics.
In nonintegrable Hamiltonian systems,
two mechanisms for instabilities leading to global diffusion in the phase space
are well known;
Arnold diffusion~\cite{Arn1964,Chi1979,Viv1984,Cin2002} and resonance overlap~\cite{Chi1979,LL1992}.
In a Hamiltonian system in general, there are resonance conditions
$\sum_i m_i\omega_i+M=0$ , where  $m_i$'s and $M$ are arbitrary integers
and $\omega_i$ is radial frequency of $i$-th element.
The conditions form resonance lines in the phase space, around which the motion is
stochastic, giving rise to a layer, while the interwoven resonance layers form so-called ``Arnold web''.
Arnold diffusion is the motion along the resonance layers,
and is a universal behavior in the systems
with more than two degrees of freedom.
Resonance overlap, on the other hand, derives from
destruction of tori which divide each resonance
layers, and results in global transport in the phase space,
as has been studied in detail in 2-dimensional mappings~\cite{Chi1979,LL1992}.

In a system with many degrees of freedom in general, both the two mechanisms coexist,
and it is not so easy to separate them out, to unveil which
part is relevant to global transport in the phase space.
In this letter, we study this problem by extracting out each mechanism separately.
First, we introduce a novel
visualization procedure to detect the structure of Arnold web and resonance
overlaps in the frequency space\cite{MDE1987,Las1990,Las1993}.
With the aid of this representation,
we measure residence time distributions at each resonance layer,
to distinguish the dynamics by Arnold diffusion from the resonance overlap.
Then, to clarify relevance of Arnold diffusion and resonance overlap to
global transport in the phase space, we compute transition diagrams
in the frequency space.  Following these results,
the global diffusion coefficient in the phase space is studied.

As a simple model for Hamiltonian system with several degrees of freedom,
we have chosen Froeschl\'{e} map~\cite{Fro1971,KB1985,WLL1990,Las1993},
given by
\begin{eqnarray}
\left\{
  \begin{array}{@{\,}ll}
   p_i(n+1)=p_i(n)+K\sin(q_i(n)) + b\sin(\sum_{k=1}^2 q_k(n))\\
   q_i(n+1)=q_i(n)+p_i(n+1)
  \end{array}
\right.
\label{eq:FroeschleMap}
\end{eqnarray}
where $i=1,2$.
$q_i(n)$ is the displacement of $i$-th element and
$p_i(n)$ its conjugate momentum.
Here, $K$ represents nonlinearity of each element and
$b$ gives the coupling strength.

Froeschl\'{e} map could be taken as a coupled system consisting of standard
maps.  The standard map, each element dynamics with $b=0$, is studied as a prototype
of a Poincar\'{e} map for Hamiltonian dynamics with two degrees of freedom.
Similarly the above Froeschl\'{e} map could be regarded as a Poincar\'{e} map of
Hamiltonian system with three degrees of freedom, and provides a
prototype model for such system.

Here, the rotation numbers
\begin{equation}
 \omega_i \equiv \lim_{T \to \infty} \frac{q_i(T)-q_i(0)}{2\pi T}
  =\lim_{T \to \infty} \sum_{n=1}^{T}\frac{p_i(n)}{2\pi T}
\end{equation}
are defined to each $i$-th element.
Later we are mainly interested in time evolution in the frequency space.
Thus, we employ the local rotation numbers computed over finite time
length $T$ by
\begin{equation}
 \omega_i(jT)\equiv\sum_{n=jT}^{jT+T-1}\frac{p_i(n)}{2\pi T}.
  \label{LocalRotationNumber}
\end{equation}
In the term of rotation numbers,
the resonance condition of Froeschl\'{e} map is given by
$ m_1\omega_1 + m_2\omega_2 + M =0 $,
where  $m_i$'s and $M$ are arbitrary integers,
and the order of resonances is defined by $\sum_i|m_i|+|M|$.
The value of $T$ must be chosen large enough to assure the
convergence of each rotation number to a certain resonance, but
not too large so that transition between the resonance layers is detected.
In this letter, we choose $T$ typically as $10^3$,
but change of it within a moderate range does not influence
our results to be reported.

To visualize the resonances and Arnold web, we use the
density plot in the frequency space, as follows:
First, we compute the rotation numbers modulo 1 over a finite time
interval from trajectories to describe the structures.
Then, the distribution of the rotation number is computed as
the histogram of the local rotation numbers.
By taking bins of some size over the frequency space $[0,1)$,
every visit at each bin is counted, to compute probability density of
the local rotation numbers.

\begin{figure}[tbp]
  \begin{center}
   \includegraphics[scale=0.48]{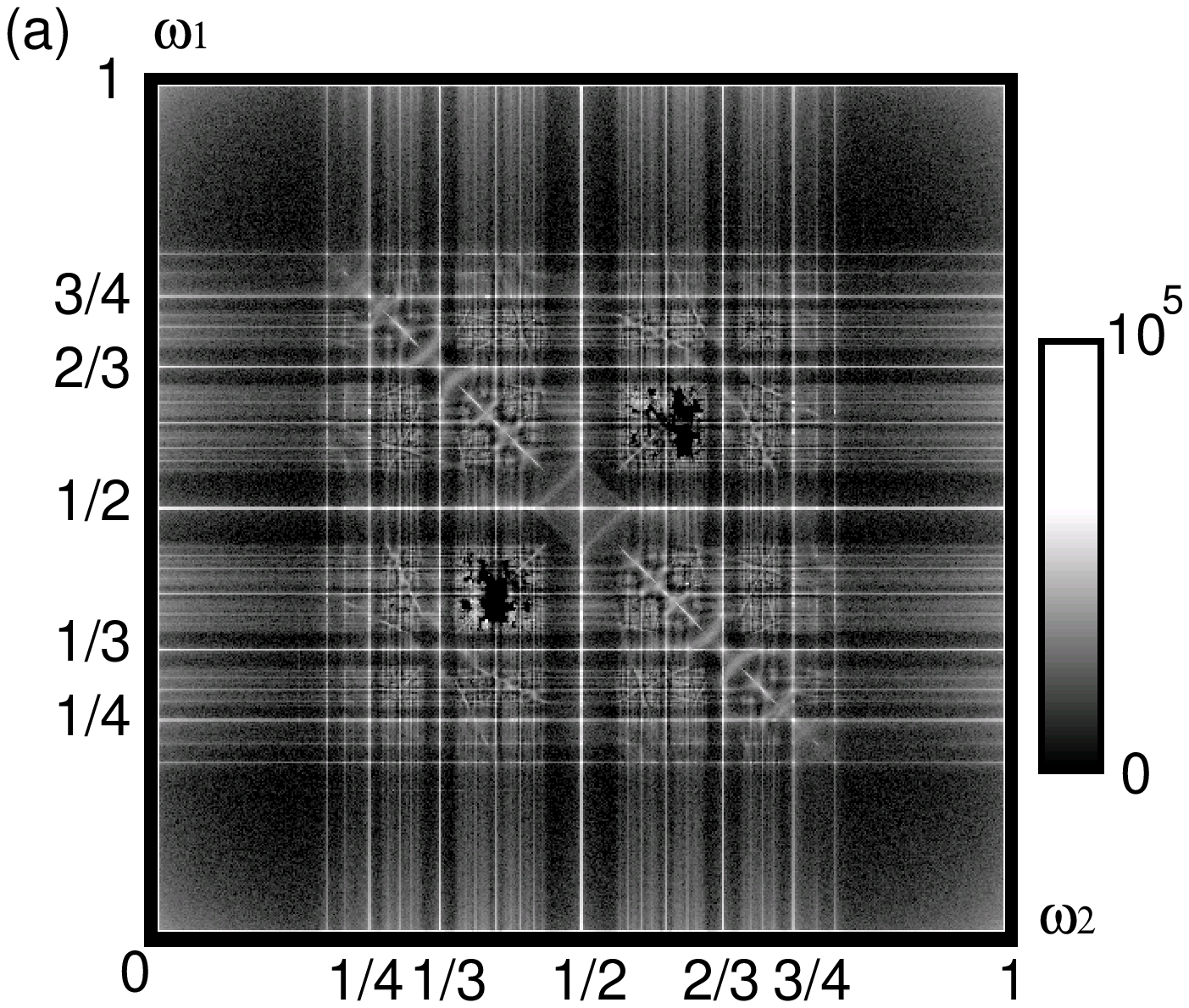}
   \includegraphics[scale=0.48]{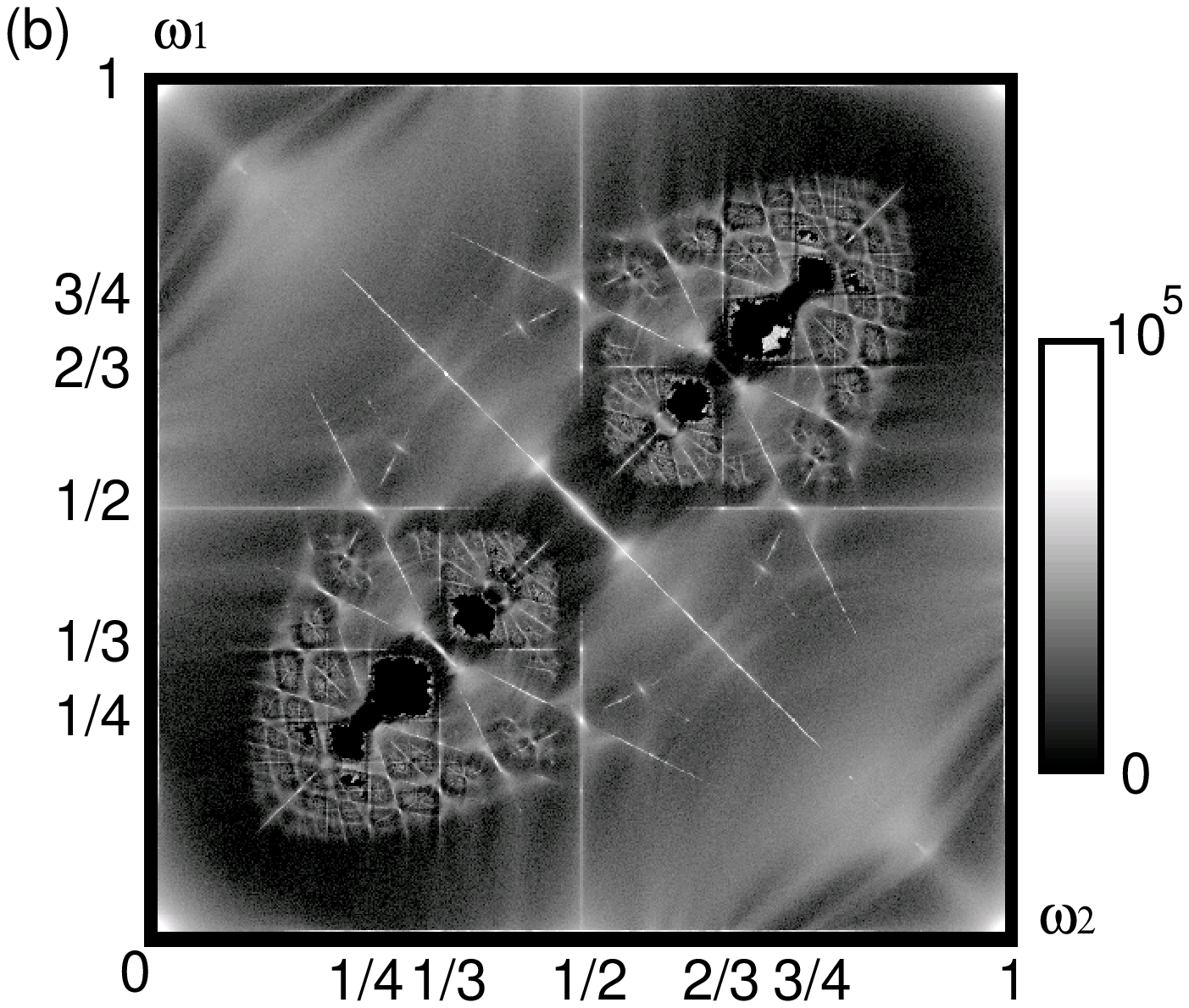}
  \end{center}
 \caption{Arnold web in the 2-dimensional frequency space.  
 The histogram of local rotation numbers, computed
 by the number of times visited at each local rotation number with
 $512\times512$ bins for $10^{10}$ iterations with $T=10^3$.
 (a) $K=0.9$, $b=0.002$ (b) $K=0.5$ and $b=0.1$.
 }
\label{Fig1.letter.eps}
\end{figure}

The density in the frequency space obtained from a single
trajectory is shown in Fig.\ref{Fig1.letter.eps}.
Vertical and horizontal resonance lines $m_i\omega_i+M=0$ mean that
each element is resonant with external force.
Among them, the lowest order resonances are $(\omega_i+1=0)$'s or $(\omega_i=1)$'s,
and resonance is higher order with larger $m_i$'s.
Lines $\sum_i m_i\omega_i+M=0$ mean that two elements are
resonant with each other, thus indicating coupling resonances.
In Fig.\ref{Fig1.letter.eps}(a) with parameters $K=0.9$ and $b=0.002$,
the coupling resonance $\omega_1+\omega_2=0$ is clearly visible,
while the other lowest-order coupling resonance $\omega_1-\omega_2=0$ is
obscure.  This is due to the coupling form of Froeschl\'{e} map.
The density structure in the frequency space with weaker nonlinearity
and stronger coupling is shown in Fig.\ref{Fig1.letter.eps}(b).
Therefore, coupling resonances of various order and resonance overlaps
are clearly visible.

With the representations of these resonance layers in mind,
we clarify quantitative differences between Arnold diffusion and
resonance overlap, by examining residence time distributions $\rho(t)$ at each
layer, given by the resonance condition $m_1\omega_1 + m_2\omega_2 + M=0$.
Since there are fluctuations in local rotation numbers
due to the finite time average, we set some threshold $W$
for each resonance condition,
so that we compute the residence time during which
$|m_1\omega_1 + m_2\omega_2 + M |< W$ is satisfied.
The threshold $W$ is chosen to be around 0.0015,
while moderate change of its value
yields almost the same distribution.

\begin{figure}[tbp]
 \begin{center}
  \includegraphics[scale=0.4]{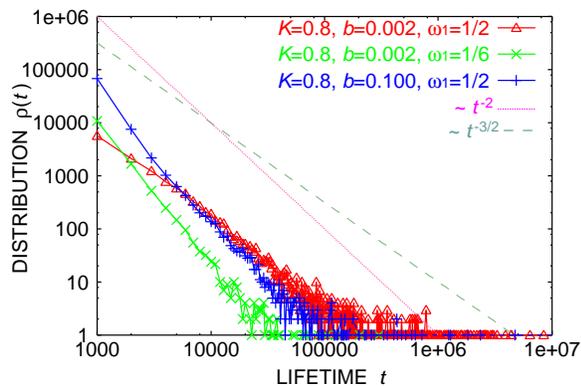}
 \end{center}
 \caption{Residence time distributions at certain resonances.
 Both dependences on the order of resonances and coupling resonances are
 shown.
 Distributions decay with a power law ($\rho(t)\sim t^{-\alpha}$).
 The exponent is $\alpha=3/2$ for lower order resonances with weak coupling ($\omega_1=1/2$, $b=0.002$),
 while $\alpha=2$ for higher order resonances ($\omega_1=1/6$, $b=0.002$) or with
 stronger coupling ($\omega_1=1/2$, $b=0.1$).
 $K=0.8$.
 }
\label{Fig2.letter.eps}
\end{figure}

First, we investigate the dependence of residence time distributions on
the order of resonances, as shown in Fig.\ref{Fig2.letter.eps}.
For all the resonances with enough residences of orbits to
get sufficient statistics, the distribution decays
with a power law ($\rho(t)\sim t^{-\alpha}$).
The exponent $\alpha$ takes the value 3/2 for lower order resonances, and
2 for higher order resonances.
The fraction of resonances with the exponent 3/2 decreases
with the increase of the coupling strength $b$ or
the nonlinearity $K$, and it is replaced by those with the exponent 2,
as shown in Fig.\ref{Fig2.letter.eps}.
These change of the exponent corresponds to the transformation of
structures in the frequency space from thin linear layer to scattered points in
2-dimensional region.  The distribution at the former has the exponent 3/2,
while the latter has the exponent 2.

This power law distribution is understood by regarding the motion at the resonance
layer as Brownian motion.
Lifetime of Brownian motion in a finite interval decays with a power law with
the exponent $\alpha=3/2$ in a 1-dimensional case and $\alpha=2$ in a
2-dimensional case.
Arnold diffusion is along the 1-dimensional resonance line, 
which is prominent at lower-order resonances when nonlinearity is weak.
In fact, the motion with the residence time distribution of the power $3/2$ is
observed for low-order resonance with weak nonlinearity.
On the other hand, overlapped resonances allow the motion across resonances which
leads to Brownian motion at a 2-dimensional region.  Indeed, the distribution with
the power $2$ is observed at higher order resonances, and is more frequently observed
with stronger nonlinearity.  Hence, one can distinguish
clearly the Arnold diffusion from resonance overlap by the
power of the residence time distribution at each resonance condition.

\begin{figure}[tbp]
  \begin{center}
   \includegraphics[scale=0.4]{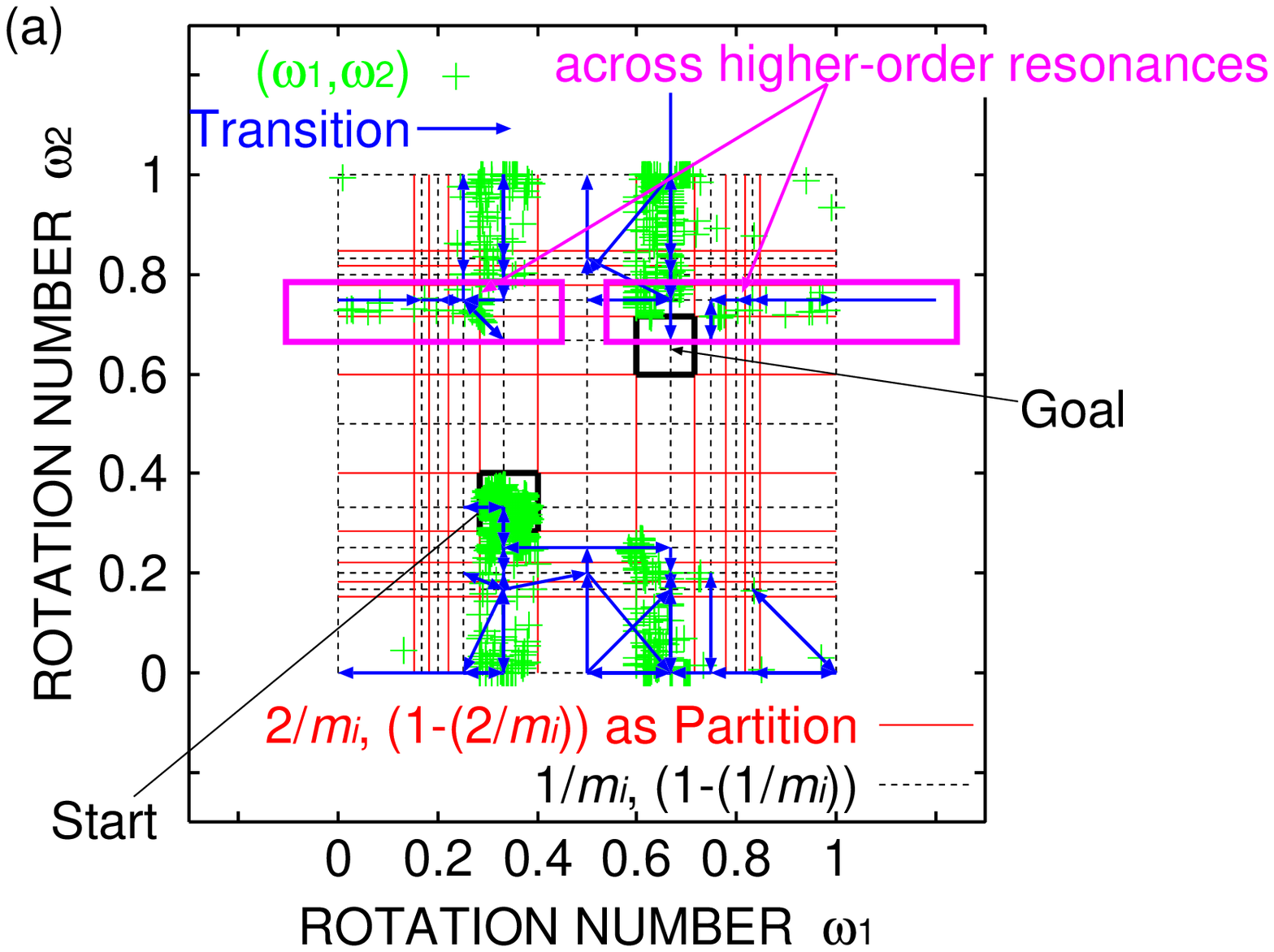}
   \includegraphics[scale=0.4]{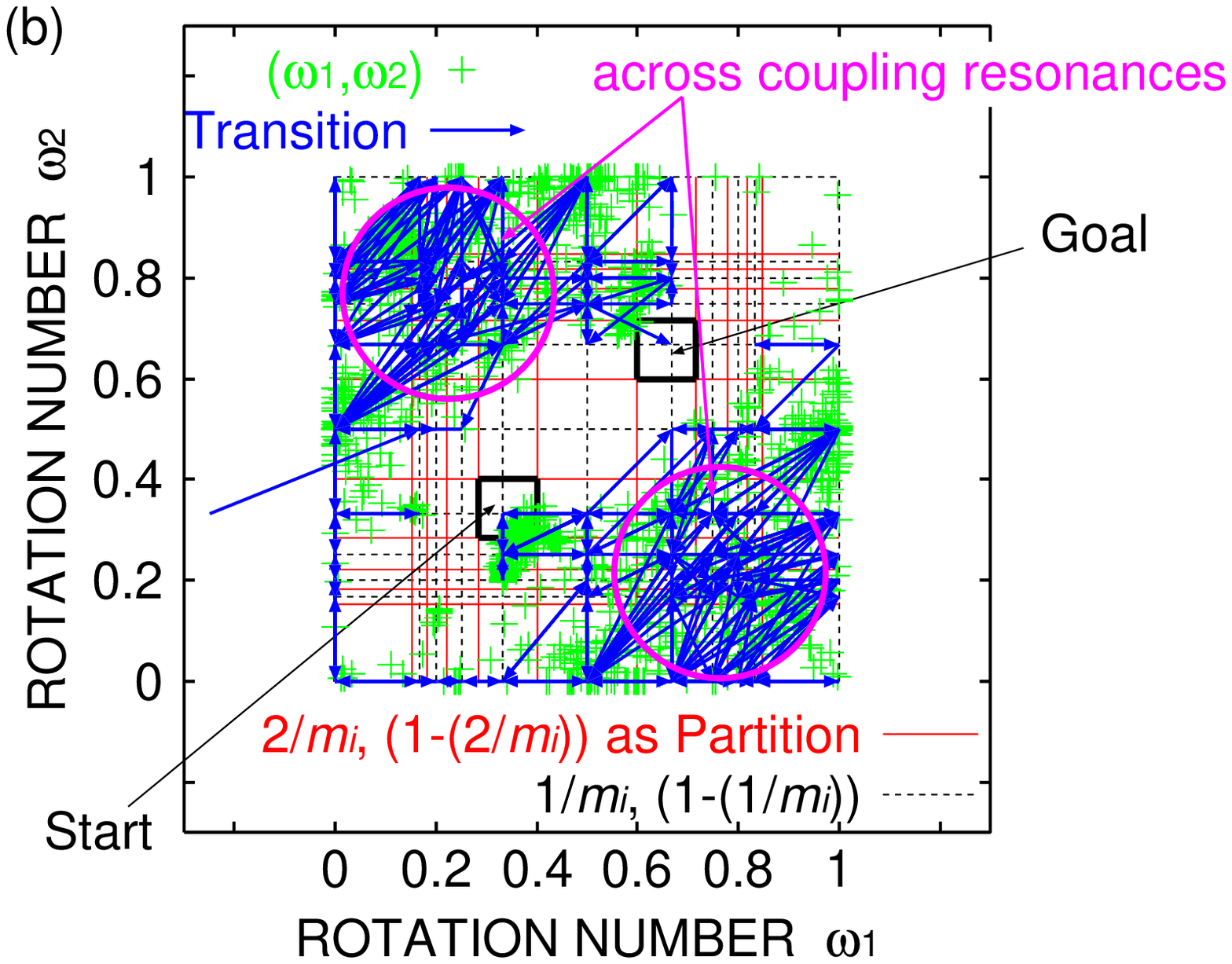}
  \end{center}
 \caption{Transition diagram in the frequency space starting from the start
 to the goal.  Plotted is the diagram for an orbit that requires
 shortest steps for the goal over randomly chosen 64 samples.
 (a) $K=0.9$, $b=0.002$. $2187 \times 10^3$ steps are required, and most 
 transitions occurs through the overlapped higher-order resonances.
 (b) $K=0.5$, $b=0.1$, $1416\times 10^3$ steps are required.
 Remarkable transitions occur through the overlapped resonances across
 the coupling resonances.
 }
 \label{Fig3.letter.eps}
\end{figure}

Now we discuss the global transport process in the phase space,
which consists of the motions across and along the resonance layers,
based on the observed geometric structures in the frequency space.
For this purpose we compute the transition diagram over resonance lines
measured in the frequency space by averaging the momentum over a finite time.
Here, we compute the transition matrix among the lower-order resonances
$\omega_i=1/m_i$ and $\omega_i=1-(1/m_i)$, where $1\leq m_i \leq 6$,
by coarse-graining the value of $\omega_i$, so that each region contains
one junction of the lower-order resonances.
As an example, we computed the transition diagram in the course of orbits
starting from the region around the golden mean torus
$\omega_i \approx (\sqrt{5}-1)/2$ and reaching that around the torus with $1-(\sqrt{5}-1)/2)$.
In other words, we study how a transition occurs from a point near one KAM torus to
another distant KAM torus, through successive transitions over resonance lines. 
We have computed the transition diagram from the start state to the
goal state, over randomly chosen 64 samples in the the frequency space.
The transition diagram depends on each sample, and that with the
shortest steps to arrive at the goal state is shown in
Fig.\ref{Fig3.letter.eps}(a)(b), corresponding to Fig.\ref{Fig1.letter.eps}(a)(b). 
Diagrams of other samples with short time steps for the destination have a similar
feature with Fig.\ref{Fig3.letter.eps}.
It contains transitions through the overlapped resonances across
the coupling resonances.  Although resonance overlaps
are not so dominant in the phase space,
the observed transition diagrams always use these resonance overlaps.

Note that motions across the overlapped resonances are faster
than Arnold diffusion along the resonances and always used in this global transition.
Moreover, regions with resonances overlap involving coupling resonances
inevitably allow for transitions to a variety of  directions.
In other words, these resonance overlaps play a role as highly connected
nodes, i.e., ``hubs''\cite{JTAOB2000} for the global transport.
By visiting the hub parts,
thus, the global transport is accelerated.

As the coupling strength is increased, the number of overlapped resonances that act as such hubs
increases, and they become widespread and overlapped each other.
Thus, the global diffusion is faster.

It is, then, natural to ask
whether transport actually occurs along resonances with weak coupling
and nonlinearity in the case that lower-order resonances are not
overlapped as shown in Fig.\ref{Fig1.letter.eps}.
From Fig.\ref{Fig3.letter.eps}(a),
it is not easy to answer the question.
Focusing on local rotation numbers themselves, however,
we conclude that transport occurs between the lower-order resonances, through
the overlapped higher-order
resonances that exist between the lower-order resonances.
Hence, the main transport is not due to the Arnold diffusion.

\begin{figure}[tbp]
 \begin{center}
  \includegraphics[scale=0.4]{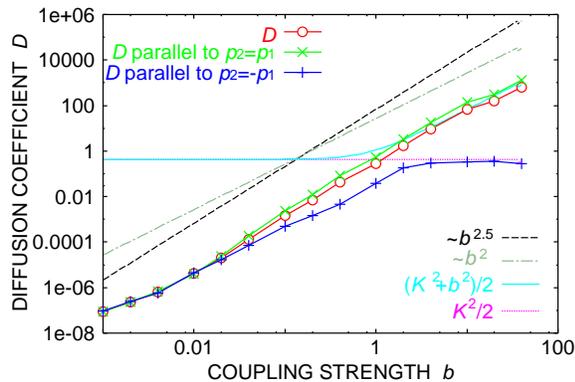}
 \end{center}
 \caption{Diffusion coefficients depending on coupling strength $b$ for $K=0.8$.
 Diffusion coefficients along certain directions are also computed.
 They are computed from 160 samples with $T=10^{8}$ in eq.(\ref{DiffusionCoefficient}).
 }
 \label{Fig4.letter.eps}
\end{figure}

So far we have shown that
the transport across the overlapped resonances is faster than that along
the resonances and existence of hubs accelerates transports.
Then, it is natural to expect that these properties of transport can influence
the macroscopic quantities which are easily obtained as diffusion coefficients.
Diffusion coefficients is defined by
\begin{equation}
D\equiv \lim_{T \to \infty} \biggl\langle\frac{1}{2}\sum_{i=1}^2\frac{(p_i(T)-p_i(0))^2}{T}\biggr\rangle,
\label{DiffusionCoefficient}
\end{equation}
where $\langle\cdot\rangle$ represents sample average.
Dependence of $D$ on the coupling
strength is shown in Fig.\ref{Fig4.letter.eps}.

Here, if we assume uniform stochasticity with no structures in the phase
space, $D=(K^{2}+b^{2})/2$ is obtained by the random phase
approximation~\cite{Chi1979}.
In general, increase with $b^2$ could be expected by replacing the
coupling term by random process as expected by diffusion.
In contrast, around $0.01\leq b \leq 1$,
the data in Fig.\ref{Fig4.letter.eps} are fitted by the form $D\sim b^{\beta}$
with $\beta=2.5$, showing a clear deviation from the above forms.
Indeed, the transport across the coupling resonance at hubs as in
Fig.\ref{Fig3.letter.eps}(b) is dominant here.
It is expected that this power law  $\beta=2.5$ could
reflect the increase of the fraction of such coupling resonances forming hubs
with the coupling $b$.
The exponent $\beta$ here decreases with $K$.
As $K$ is increased from $0.5$ to $0.9$, $\beta$ decreases from $3.5$ to $2.2$.
This dependency suggests that the increase of hub coupling resonances
is more relevant as the nonlinearity is weaker.

The relevance of diffusion across the coupling resonance is
also detected by computing the diffusion
parallel to $p_2=p_1$ (across $\omega_1+\omega_2=0$) and
to $p_2=-p_1$ (along $\omega_1+\omega_2=0$), separately~\cite{footnote}.
As the former measures the diffusion transversal to the resonance line $\omega_1+\omega_2=0$,
a main source for the difference between the two diffusion coefficients
is the motion crossing the lowest-order coupling resonance, $\omega_1+\omega_2=0$.
As shown, the diffusion coefficient parallel to $p_2=p_1$ is much larger than the other, and
the diffusion is anisotropic in this sense.
The difference is prominent for $b > 0.01$,
where the transition diagram as in Fig.\ref{Fig3.letter.eps}(b) is
observed, showing the dominance of the motion
across the hub coupling resonance.  This again demonstrates the
relevance of the motion across the resonance layer.

So far, several estimates of diffusion coefficients have been proposed
beyond the simple random phase approximation,
as given by the stochastic pump or three resonance model and their
extensions~\cite{WLL1990,CV}.
These, however, assume only the diffusion along the resonances,
while our results, in contrast, exhibit the importance of resonance
overlap to the global diffusion.

In summary,
intermingled structure of Arnold web and resonance overlaps
are visualized by introducing the representation in  the frequency space.
It is shown that the motion across the resonances is faster than the
motion along the overlapped resonances.
By examining  the transition over resonance layers,
global diffusion in the phase space is found to be
mainly governed by diffusion across the overlapped resonances.
The global transport in the phase space is accelerated
through overlapped resonances involving coupling resonances, forming a
hub in the transition diagram.
There, the global diffusion in the phase space is governed mainly by
the motion across the resonance layers, rather than the Arnold diffusion
along the layer. Now it is important to estimate the resonance overlap
in conjunction with the Arnold diffusion, to study global transport process.

In a system with more degrees of freedom,
Arnold webs may be intermingled, and form a complicated structure, 
so to say, ``Arnold spaghetti''~\cite{Kan2002}. In such case, diffusion across the
resonance layers through resonance overlaps is more dominant, where
network of hub transport regions should be unveiled,
to understand conditions for the realization of thermodynamic behavior.

This work was partially supported by a Grant-in-Aid for Scientific Research
from the Ministry of Education, Science, and Culture of Japan.


\begin{thebibliography}{15}
\bibitem{footnote}
The map(1) is written as
$ P_1(n+1)=P_1(n)+\sqrt2 K\sin(Q_1(n)/\sqrt2)\cos(Q_2(n)/\sqrt2)+\sqrt2b\sin(\sqrt2Q_1(n))$,
$ P_2(n+1)=P_2(n)+\sqrt2 K\cos(Q_1(n)/\sqrt2)\sin(Q_2(n)/\sqrt2)$,
$ Q_i(n+1)=Q_i(n)+P_i(n+1)$ ($i=1,2$)
by transforming the variable
$P_1\equiv(p_1+p_2)/\sqrt2$,$P_2\equiv(-p_1+p_2)/\sqrt2$,
$Q_1\equiv(q_1+q_2)/\sqrt2$,$Q_2\equiv(-q_1+q_2)/\sqrt2$.
Thus, only the dynamics of $P_1$ (the motion parallel to $p_2=p_1$)
includes the coupling term by $b$, while that of $P_2$ (motion  parallel
to $p_2=-p_1$) is only triggered by random phase, without
direct coupling term.
\end{thebibliography}
\end{document}